\documentclass[onecolumn,superscriptaddress,
nofootinbib,
 amsmath,amssymb,
 aps,
]{revtex4-2}

\usepackage{graphicx}
\usepackage{color}
\usepackage{latexsym}
\usepackage{slashed}

\usepackage{amssymb}

\usepackage[hidelinks]{hyperref}

\usepackage{mathrsfs}

\newcommand{\be}{\begin{eqnarray}}
\newcommand{\ee}{\end{eqnarray}}

\newcommand{\bea}{\begin{eqnarray}}
\newcommand{\eea}{\end{eqnarray}}

\newcommand{\beq}{\begin{equation}}
\newcommand{\eeq}{\end{equation}}

\renewcommand{\thesection}{\arabic{section}}

\numberwithin{equation}{section}

\let\oldsection\section
\renewcommand{\section}{
  \renewcommand{\theequation}{\thesection.\arabic{equation}}
  \oldsection}

\begin{document}


\title{Gravitational Aharonov-Bohm effect: theory and experiment}

\author{Zhongyou Mo}
\email[]{mozhy7@mail.sysu.edu.cn}
\affiliation{School of Science, Sun Yat-Sen University, Shenzhen, 518107, China}

\author{Leonardo Modesto}
\email[]{leonardo.modesto@unica.it}
\affiliation{Dipartimento di Fisica, Universit\`a di Cagliari, Cittadella Universitaria, 09042 Monserrato, Italy}
\affiliation{I.N.F.N, Sezione di Cagliari, Cittadella Universitaria, 09042 Monserrato, Italy}

\date{}

\begin{abstract}
We study the quantum interference of two paths that enclose a slowly rotating, infinitely long, and infinitely thin cylindrical shell, using the spacetime metric induced by the gravitational source. We show that the rotation of the shell affects the interference pattern even though it does not contribute to the classical motion of particles propagating outside the shell. Furthermore, we find that the gravitomagnetic field is non-vanishing only in the region inside the shell, yet it contributes to the interference taking place entirely in the exterior region. This phenomenon is therefore a gravitational analogue of the Aharonov--Bohm effect, in which a field confined to an inaccessible region influences the quantum phase of particles propagating in a field-free region.
\end{abstract}


\maketitle
\tableofcontents

\section{Introduction}

It is well known that the motion of a charged particle in an electromagnetic field is formulated in fundamentally different ways in classical electrodynamics and in quantum mechanics. In the classical framework, the particle is treated as a point-like object following a well-defined trajectory, determined by the Lorentz force equation. This equation depends explicitly on the electric and magnetic fields, which are the physically measurable quantities in classical electromagnetism. In quantum mechanics, however, the situation changes substantially. There, the state of the particle is described by a wave function satisfying a wave equation—such as the Schrödinger equation in the non-relativistic regime—and this equation is formulated directly in terms of the electromagnetic potentials, rather than the fields themselves. Although in classical physics the potentials are typically regarded as purely mathematical conveniences without direct physical meaning, Aharonov and Bohm showed in their renowned work~\cite{Aharonov:1959fk} that this is no longer true in the quantum domain. In particular, they showed that the electromagnetic potential can produce observable physical effects on the quantum particle, even in regions where the electric and magnetic fields vanish. This result reveals a deep and surprising difference between classical and quantum descriptions of electromagnetic interactions.

In the thought experiment proposed by Aharonov and Bohm, two parts of a coherent electron beam are made to interfere after traveling along paths that enclose a cylindrical solenoid. The solenoid is designed in such a way that the magnetic field is entirely confined within its interior, while the region outside—where the electrons propagate—is completely free of magnetic field. Nevertheless, the experiment shows that the presence of the confined magnetic field produces a measurable shift in the interference pattern, corresponding to a phase difference between the two parts of the electron beam. This result is striking because it demonstrates that the electrons are physically influenced by the electromagnetic potential even in regions where the magnetic field vanishes, thereby confirming that the potential has a direct physical significance in quantum mechanics.

Beyond its implications for the role of electromagnetic potentials, the Aharonov–Bohm effect (AB effect) also provides a clear and important manifestation of the nonlocal character of quantum mechanics. Indeed, the phase shift acquired by the electrons depends on the magnetic flux enclosed by their paths, even though the electrons never enter the region where the magnetic field is present. This suggests that quantum particles can be sensitive to global properties of the electromagnetic field configuration, rather than only to local forces, in sharp contrast with the classical intuition.

Shortly after the theoretical proposal by Aharonov and Bohm, the magnetic version of the effect was experimentally confirmed by Chambers in 1960~\cite{Chambers:1960xlk}, providing direct evidence for the predicted phase shift. Since then, the AB effect has been extensively developed and has found applications and generalizations in numerous areas of physics, ranging from condensed matter systems to gauge theories and mesoscopic physics. Detailed and comprehensive treatments of both the theoretical foundations and the experimental realizations of the effect can be found, respectively, in Ref.~\cite{Peshkin:1989zz} and Ref.~\cite{Tonomura:1989zz}.

A recent experiment reported in Ref.~\cite{Overstreet:2021hea} has drawn renewed attention to the gravitational analogue of the Aharonov--Bohm effect. In that work, the authors claimed to have observed a gravitational Aharonov--Bohm effect in a laboratory setting (for a concise introduction to the experiment, see also Ref.~\cite{Roura:2021fvd}). Specifically, the experiment measured a gravity-induced phase shift for atoms, given by
\begin{equation}
\phi=\frac{m}{\hbar}\int \bigl[
V(x_1,t)-V(x_2,t) \bigr]dt \, ,
\label{V1minusV2}
\end{equation}
where $V(x_i, t)$ denotes the gravitational potential evaluated at the position $x_i$ at time $t$, and the gravitational source was realized by means of a ring-shaped mass distribution. It is important to observe that the phase \eqref{V1minusV2} is formally analogous to the phase appearing in the electric version of the Aharonov--Bohm effect~\cite{Aharonov:1959fk}. However, in the experiment of Ref.~\cite{Overstreet:2021hea}, the measurement was not performed in an ideal region where the gravitational force vanishes identically. A proposal for an experiment carried out in a region with zero gravitational force, together with a theoretical demonstration that a phase shift can nevertheless be induced by nearby masses, was put forward earlier in Ref.~\cite{Hohensee:2011yt}.

Although the experiment of Ref.~\cite{Overstreet:2021hea} represents a major step forward, it is worth emphasizing that the phase \eqref{V1minusV2} does not capture the full physical content of the gravitational Aharonov--Bohm effect. In the framework of gravitoelectromagnetism (GEM), a weak gravitational field can be described in terms of both a scalar potential and a vector potential, in close analogy with classical electromagnetism~\cite{Mashhoon:2003ax,Gallerati:2022pgh,Harris1991}. The phase \eqref{V1minusV2} manifestly contains only the contribution of the scalar potential. For a general weak gravitational field, however, the vector potential is also expected to contribute and should be taken into account~\cite{Ho1994}. In this direction, a gravitational Aharonov--Bohm effect induced purely by the vector potential was investigated in Ref.~\cite{Chiao:2013ht}.

Beyond the GEM formulation, alternative approaches to the gravitational Aharonov--Bohm effect have also been developed in the literature. A conceptually different definition of the effect appears, for instance, in Refs.~\cite{Dowker,Ford,Audretsch:1982ux}, where it was shown that particles propagating in a region of vanishing curvature can nonetheless be influenced by the non-zero Riemann tensor of an inaccessible region. This formulation highlights the deep geometric nature of the gravitational Aharonov--Bohm effect and its connection with the underlying structure of spacetime.

In the present paper, we restrict our attention to weak gravitational fields and adopt the GEM approach to investigate the gravitational Aharonov--Bohm effect. In Sec.~\ref{GEM}, we briefly review the GEM formalism and its derivation from the spacetime metric. In Sec.~\ref{phase}, building on the GEM framework, we derive the Schr\"odinger equation and the corresponding gravitational phase for a non-relativistic particle. In Sec.~\ref{GAB effect}, we apply these results to discuss a concrete gravitational Aharonov--Bohm interference experiment. Throughout the paper, we use the metric signature $(-,+,+,+)$.

\section{Gravitoelectromagnetism\label{GEM}}
In this section we explicitly show that in the weak field approximation and large $c$ (speed of light) limit the Einstein's equations of motion take the formal expression of Maxwell's equations for gravity. 

In the weak gravity approximation, we can decompose the metric as follows,  
\be
g_{\mu\nu}=\eta_{\mu\nu} +h_{\mu\nu} \, ,
\ee
where $\eta_{\mu\nu}={\rm diag}(-1,1,1,1)$ is the Minkowski metric and $h_{\mu\nu}$ is a small perturbation satisfying $|h_{\mu\nu}|\ll 1$. In this approximation, the indexes are raised and lowed by the metric 
$\eta_{\mu\nu}$. Hence, using the gauge condition~\cite{Misner1973}
\be
 \partial^\mu \bar{h}_{\mu\nu}=0 \, ,
\label{gauge condition} 
\qquad 
 \bar{h}_{\mu\nu} \equiv h_{\mu\nu}-\frac{1}{2} {h^\alpha}_\alpha \eta_{\mu\nu}\, , 
\ee
the Einstein field equations simplify to the following wave equation \cite{Mashhoon:2003ax}, 
\be
\square \bar{h}_{\mu\nu} =-\frac{16\pi G}{c^4} T_{\mu\nu} \, ,
\label{box barH}
\ee
which has the following retard solution, 
\be
\bar{h}_{\mu\nu}
=\frac{4G}{c^4} \int \frac{T_{\mu\nu} (ct -|\vec{x}-\vec{x}'|, \vec{x}')}{|\vec{x}-\vec{x}'|} d^3 x' \, .
\label{source integration}
\ee
The energy momentum tensor, making explicit the dependence on the speed of the light, reads:
\be
T_{00} = \rho c^2 \, , \quad T_{0j} =-c\rho u^j \, , \quad \mbox{and} \quad  \quad T_{ij} = \rho u_i u_j +p\delta_{ij} \, . 
\label{Tc}
\ee
Therefore, according to (\ref{Tc}), the solution (\ref{source integration}) implies the following dependence of the perturbations of the metric on the speed of the light, 
\be
\bar{h}_{00}=O(c^{-2}) \, ,  \quad \bar{h}_{j0}=O(c^{-3})\, ,  \quad \bar{h}_{ij}=O(c^{-4}) \, . 
\label{hsub}
\ee
If all the terms of order $O(c^{-4})$ are neglected, the spacetime metric can be approximated by~\cite{Mashhoon:2003ax}:
\be
ds^2= -c^2\Bigl(1-\frac{1}{2} \bar{h}_{00} \Bigr) dt^2 +2c \bar{h}_{0i} dx^i dt +\Bigl(1+\frac{1}{2} \bar{h}_{00} \Bigr) \delta_{ij} dx^i dx^j \, ,
\label{ds bar}
\ee
namely $\bar{h}_{ij}$ does not appear in \eqref{ds bar} because $\bar{h}_{ij}$ is negligible compared to $\bar{h}_{0\mu}$ according to (\ref{hsub}). Therefore, in the above approximation, a gravitoelectromagnetic theory will be defined for $\bar{h}_{\mu 0}$.

To make explicit the analogy with electromagnetism, let us rewrite \eqref{box barH} (in which \eqref{gauge condition} has been used) as follows \cite{Gallerati:2022pgh}, 
\be
\frac{c^2}{2} \partial^\rho (\partial_\nu \bar{h}_{\rho \mu} -\partial_\rho \bar{h}_{\nu \mu}) =\frac{8\pi G}{c^2} T_{\mu\nu} \, , \quad 
 \partial^\mu \bar{h}_{\mu\nu} = 0 \, .
\label{Einstein equation2}
\ee
Notice that the first term in the above equations is actually zero in the Lorentz's gauge.
Hence, (\ref{Einstein equation2}) and (\ref{box barH}) are equivalent. However, such redundant form of the equations of motion together with the gauge condition will be useful shortly. Since $\bar{h}_{ij}$ is negligible in comparisons to the other components in (\ref{hsub}), we do not take into account of the $ij$ components of equation (\ref{Einstein equation2}), but we only focus on the $(\mu = 0, \nu)$ components of the first equation in (\ref{Einstein equation2}) and the $(\nu = 0)$ component of the second equation, namely the zero component of the gauge condition. Hence from (\ref{Einstein equation2}) we get:
\be
\frac{c^2}{2} \partial^\rho (\partial_\nu \bar{h}_{\rho 0} -\partial_\rho \bar{h}_{\nu 0}) =\frac{8\pi G}{c^2} T_{0 \nu} \, , \quad 
\partial^\mu \bar{h}_{\mu 0 } = 0 \, .
\label{Einstein equation3}
\ee
Then, following \cite{Gallerati:2022pgh}, we  define the gravitoelectromagnetic potential as:
\be
A_\nu := \frac{1}{2}c^2\bar{h}_{\nu 0} \, ,
\label{Anu}
\ee
and the equations (\ref{Einstein equation3}) turn into: 
\be
\frac{2}{c^2} \frac{c^2}{2} \partial^\rho (\partial_\nu A_{\rho} -\partial_\rho A_{\nu}) =\frac{8\pi G}{c^2} T_{0 \nu} \, , \quad 
 \partial^\mu A_{\mu} = 0 \, .
\label{Einstein equation4}
\ee
We can also introduce the gravitoelectromagnetic field strength, namely 
\be
F_{\nu\rho}:=\partial_\nu A_\rho -\partial_\rho A_\nu \, ,
\qquad
E_i :=-F_{0i} \, ,
\qquad
B_i :=\frac{1}{2}{\varepsilon_i}^{jk}F_{jk} \, ,
\label{G0 nu rho}
\ee
(where ${\varepsilon_i}^{jk}$ is the $3$-dimensional Levi-Civita symbol, e.g.,  ${\varepsilon_1}^{23}=-{\varepsilon_1}^{32}=1$)
and the following gravitational current, 
\be
J_\mu :=T_{0\mu} \, ,
\ee
so that the Einstein's equations (\ref{box barH}), in the Lorentz's gauge (\ref{gauge condition}), in the weak field limit $|h_{\mu\nu}| \ll 1$, and at the order $O(1/c^4)$ take the same form of the Maxwell's equations, namely
\be
\partial^\rho F_{\nu \rho} =\frac{8\pi G}{c^2} J_\nu \, , \quad  \partial^\mu A_{\mu} = 0 \, . 
\label{current equation}
\ee
Let us now focus on the symmetry of the theory (\ref{current equation}). Despite the potential $A_\mu$ in the equations of motion (EoM) (\ref{current equation}) satisfies the gauge condition $\partial^\mu A_{\mu} = 0$, the EoM (\ref{current equation}) and the gauge condition $\partial^\mu A_{\mu} = 0$ both still enjoy the following residual $U(1)$ invariance (similarly to electromagnetism), 
\be
A'_\mu=A_\mu +\partial_\mu \bar{\Lambda} \, , \quad \mbox{with} \quad \Box \bar{\Lambda} = 0 \, . 
\label{transformation A}
\ee
The latter harmonic condition on $\bar{\Lambda}$  is necessary in order to maintain the condition $\partial^\mu A_{\mu} = 0$. 
Indeed, 
\be
\partial^\mu A_{\mu}^\prime  = \partial^\mu A_{\mu} +  \Box \bar{\Lambda} \, , 
\ee
and $\bar{\Lambda}$ has to satisfy $\Box \bar{\Lambda} = 0$ to preserve the gauge condition 
$\partial^\mu A_{\mu}^\prime = 0$. Finally, the number of physical degrees of freedom, taking into account of the gauge condition $\partial^\mu A_{\mu} = 0$ and the residual gauge invariance (\ref{transformation A}), is two exactly like in electromagnetism. 

Having checked that the EoM, the gauge condition, and the residual gauge invariance are exactly the same met in electromagnetism, we can now relax the gauge condition 
$\partial^\mu A_{\mu} = 0$ and replace the two equations (\ref{current equation})
with only the first one, 
\be
\partial^\rho F_{\nu \rho} =\frac{8\pi G}{c^2} J_\nu \, , 
\label{current equation2}
\ee
but now enjoying the full $U(1)$ gauge invariance,
\be 
A'_\mu=A_\mu +\partial_\mu {\Lambda} \, ,
\ee
without any restrictions on the local function $\Lambda$, and Lorentz invariance. Since there are no restrictions on $\Lambda$, the EoM (\ref{current equation2}) are equivalent to the two equations (\ref{current equation}). 

For completeness, according to the above definitions (\ref{G0 nu rho}), the analog of the Maxwell equations for gravity read (see Ref.~\cite{Gallerati:2022pgh} for a complete list of papers in which the model here presented has been derived)\footnote{The fields $\vec{E}$ and $\vec{B}$ in our equations have the same dimension of $LT^{-2}$ contrary to the Ref.~\cite{Gallerati:2022pgh}.}:
\bea
\nabla\cdot \vec{E} &=& -8\pi G\rho \, , 
\nonumber 
\\
\nabla\times \vec{E} &=&-\partial_0 \vec{B} \, ,
\nonumber 
\\
\nabla\cdot \vec{B} &=&0 \, ,
\nonumber 
\\
\nabla\times \vec{B} &=&\partial_0\vec{E} +\frac{8\pi G}{c^2}
\vec{J}  \, , 
\label{Maxwell like}
\eea
where $\rho=T_{00}/c^2$ and $J^i=T_{0i}$.

\section{Gravitational phase\label{phase}}
The weak field and low speed formulation of gravity summarized in the previous section are crucial in order to make the AB effect for gravity very similar to the electromagnetic counterpart. 
In this section, we derive the classical Lagrangian and the Schr\"odinger equation for a point-like particle 
consistently with the approximations implemented in Einstein's gravity. In particular, the latter equation will be the cornerstone for deriving the AB effect due to gravity. 

\subsection{Lagrangian and Hamiltonian\label{dynamics}}

As shown in the textbook of quantum mechanics~\cite{Greiner2001}, one usually needs to find an equation between the Hamiltonian and the canonical momentum then replace them by the corresponding operators. Let us firstly derive the Lagrangian for a massive particle. The Lagrangian is related to the proper time $\tau$
by means of the action, 
\be
S =\int L dt
= - m c^2 \int  d \tau \, .
\label{S and L}
\ee
Therefore, similarly to the derivation in Ref.~\cite{Stodolsky:1978ks}, we expand the line element at the linear level in $h_{\mu\nu}$ and we parametrize the world line in the proper time respect to the Minkowski metric, namely
\bea
c d\tau
&=&c d\tau_0 \sqrt{1-\frac{h_{\mu\nu}}{c^2} \frac{dx^\mu}{d\tau_0} \frac{dx^\nu}{d\tau_0} },
\nonumber\\
&\approx& c d\tau_0 
\Bigl[1-\frac{1}{2} h_{00} \Bigl(\frac{dt}{d\tau_0} \Bigr)^2
-h_{j0} \frac{dt}{d\tau_0} \frac{dx^j}{c d\tau_0} 
-\frac{1}{2} h_{ij} \frac{dx^i}{c d\tau_0} \frac{dx^j}{c d\tau_0} 
\Bigr]
\nonumber\\
&=& c dt 
\Bigl[\frac{d\tau_0}{dt}
-\frac{1}{2} h_{00} \frac{dt}{d\tau_0}
-h_{j0} \frac{dx^j}{c d\tau_0} 
-\frac{1}{2} h_{ij} \frac{d\tau_0}{dt} \frac{dx^i}{c d\tau_0} \frac{dx^j}{c d\tau_0} 
\Bigr],
\label{cdtau}
\eea
where $(c d\tau_0)^2=-\eta_{\mu\nu} dx^\mu dx^\nu$. From the first to the second step we expanded the square root at the first order in the gravitational perturbation, and neglected the metric perturbations $h_{\mu\nu}$ higher than $O(c^{-3})$ according to \eqref{hsub}. Moreover, we implemented the relations (\ref{gauge condition}):
\be
h_{00}=\frac{1}{2}\bar{h}_{00},
\qquad
h_{0j}=\bar{h}_{0j},
\qquad
h_{ij}=\frac{1}{2}\bar{h}_{00} \, \delta_{ij} \, .
\label{hhh}
\ee 
Plugging \eqref{hhh} into \eqref{cdtau}, we get
\be
c d\tau
\approx
 c dt
\Bigl[\frac{d\tau_0}{dt}
-\frac{1}{4} \bar{h}_{00} \frac{dt}{d\tau_0}\Bigl(1+ \delta_{ij}
\frac{dx^i}{c dt} \frac{dx^j}{c dt} \Bigr)
-\bar{h}_{j0} \frac{dt}{d\tau_0} \frac{dx^j}{c dt}
\Bigr] \, .
\label{cdtau2}
\ee
Comparing \eqref{cdtau2} with \eqref{S and L}, we get the Lagrangian
\bea
L
&=&
-mc^2 
\Bigl[\frac{d\tau_0}{dt}
-\frac{1}{4} \bar{h}_{00} \frac{dt}{d\tau_0}\Bigl(1+ \delta_{ij}
\frac{dx^i}{c dt} \frac{dx^j}{c dt} \Bigr)
-\bar{h}_{j0} \frac{dt}{d\tau_0} \frac{dx^j}{c dt}
\Bigr]
\nonumber\\
&=&
{
-mc^2 
\Bigl\{
\Bigl(1-\frac{v^2}{c^2}\Bigr)^{1/2}
-\Bigl(1-\frac{v^2}{c^2}\Bigr)^{-1/2}\Bigl[
\frac{1}{4} \bar{h}_{00}  \Bigl(1+ 
\frac{v^2}{c^2} \Bigr)
+\bar{h}_{j0}  \frac{v^j}{c}
\Bigr]
\Bigr\}
 \, ,
 }
\label{Lagrangian}
\eea
which coincides with (1.10) of Ref.~\cite{Mashhoon:2003ax}, where $v^i = dx^i/dt$. 

{As for non-relativistic particles, we neglect the terms of the order higher than $O(v^2/c^2)$ and the terms $v^2/c^2$ that multiply 
$\bar{h}_{\mu\nu}$ in \eqref{Lagrangian}. Moreover, we discard the rest energy. Finally, the Lagrangian reades:}
\bea
L\approx \frac{1}{2} m v^2
+\frac{1}{4}mc^2\bar{h}_{00} 
+mc \bar{h}_{j0} v^j.
\label{Lagrangian6}
\eea
In analogy with electromagnetism, we introduce the vector $K_\mu$ defined by:
\be
K_0=\frac{1}{4} c^2 \bar{h}_{00} \, ,
\qquad
K_j=c^2 \bar{h}_{j0} \, ,
\label{K0Kj}
\ee
thus, we can formally recast the Lagrangian \eqref{Lagrangian6} in the following form,
\be
L=\frac{1}{2} m v^2
+mK_0
+\frac{m}{c} K_j v^j \, ,
\label{Lagrangian7}
\ee
while the action reads:
\be
S=\int \Bigl( \frac{1}{2} m v^2 dt
+\frac{m}{c} K_\mu dx^\mu\Bigr)  \, .
\label{action}
\ee
However, it deserves to be notice that the vector $K_\mu$ is not a Lorentz covariant four-vector, contrary to  $A_\mu$ defined in \eqref{Anu}. 
In particular, comparing \eqref{K0Kj} and \eqref{Anu}, we get the following relation between the components of $K_\mu$ and $A_\mu$, 
\be
K_0=\frac{1}{2} A_0 \, ,
\qquad
K_j=2 A_j \, .
\label{relation KA}
\ee
Hence, we introduced (\ref{K0Kj}) for convenience in analogy with the action of a point particle in presence of electromagnetic fields, but we do not need to require Lorentz invariance. Indeed, since later we will consider the Newtonian dynamics, or actually the quantum Schroedinger's dynamics, we do not need to worry about the manifest Lorentz covariance of the action (\ref{action}). 

Given the Lagrangian (\ref{Lagrangian7}) and following the book~\cite{Landau:1975pou}, we immediately get the canonical momentum and the Hamiltonian, namely:
\bea
\vec{P}&=& \frac{\partial L}{\partial \vec{v} }
=m\vec{v} +\frac{m}{c}\vec{K} \, ,
\label{canonical momentum}
\\
H&=&\vec{v}\cdot\vec{P} -L
=\frac{1}{2}mv^2 -mK_0 \, .
\label{Hamiltonian}
\eea
Comparing \eqref{Hamiltonian} with \eqref{canonical momentum}, we get the relation:
\be
\frac{1}{2m}\Bigl(\vec{P}- \frac{m}{c}\vec{K}\Bigr)^2
=H+mK_0 \, .
\label{PKHK}
\ee
It deserves to be mentioned that in Ref.~\cite{DeWitt:1966yi}, the author derived the same equation  \eqref{PKHK} but in the presence of electromagnetic fields and gravity, and it reduces to \eqref{PKHK} when the gravity is only contained (one can check this consistence by substituting \eqref{K0Kj} and \eqref{hhh} into \eqref{PKHK}).


\subsection{Schr\"odinger equation and gravitational phase\label{Schrodinger}}
In this section, we show that the Schr\"odinger equation corresponding to \eqref{PKHK} leads to a path-dependent phase~\cite{Dowker:1967zz}. Replacing the Hamiltonian and  the canonical momentum in \eqref{PKHK} by the corresponding quantum operators, we get the Schr\"odinger equation in the present of a weak gravitational field, 
\be
\frac{1}{2m}\Bigl(-i \hbar \nabla -\frac{m}{c}\vec{K}\Bigr)^2 \Psi
=(i\hbar \partial_t +mK_0)\Psi \, .
\label{Schrodinger equation}
\ee
Hence, according to~\cite{Ho1994}, we define the following ``gauge covariant" derivative\footnote{We will see in Sec.~\ref{gauge properties} the gauge transformation associated to \eqref{gauge derivative} is subtle, compared with the case of electromagnetism.}, 
\be
\mathscr{D}_\mu =\partial_\mu -\frac{im}{\hbar c} K_\mu \, ,
\label{gauge derivative}
\ee
by which the Schr\"odinger equation \eqref{Schrodinger equation} can be written in a more compact notation, 
\be
-\frac{\hbar^2}{2m}\vec{\mathscr{D}}^2 \Psi
=i\hbar c \mathscr{D}_0 \Psi \, .
\label{gravitational wave equation}
\ee
Finally, we get the following solution of the equation (\ref{gravitational wave equation}) \cite{Ho1994}:
\be
\Psi(t,\vec{x})=\Psi_0(t,\vec{x}) e^{i\frac{m}{\hbar c}\int K_\mu dx^\mu},
\label{wave function}
\ee
where $\Psi_0(t,\vec{x})$ is the free wave function satisfying the Schr\"odinger equation with $K_\mu=0$. We notice that in \eqref{wave function} there is a gravitational phase related to the gauge potential, 
\be
\phi=\frac{m}{\hbar c}\int K_\mu dx^\mu,
\label{gravitational phase}
\ee 
which is consistent with (11) in Ref.~\cite{Dowker:1967zz}, considering \eqref{K0Kj} and \eqref{hhh}.

\subsection{Gauge properties\label{gauge properties}}

We here briefly discuss the gauge invariance of equation~\eqref{gravitational wave equation}. One can see that the wave equation \eqref{gravitational wave equation} is invariant under the transformations: 
\be 
K'_\mu=K_\mu +\partial_\mu g(x) \quad \mbox{and} \quad 
\Psi'=\Psi e^{\frac{im}{\hbar c} g(x)} \, , 
\ee
 where $g(x)$ is a general function of the spacetime point $x$. However, the gravitoelectromagnetic fields are not invariant under the former transformation, but under (\ref{transformation A}). Therefore, the correct transformations for the components of $K_\mu$ are:
\be
K'_0=K_0 +\partial_0 g(x) \, ,
\qquad
K'_j=K_j +4\partial_j g(x) \, ,
\label{K0Kj transform}
\ee
with $g(x)=\Lambda(x)/2$. Indeed, replacing $g(x)=\Lambda(x)/2$ in (\ref{K0Kj transform}) and using the identifications (\ref{relation KA}) we get (\ref{transformation A}). 

Since the transformation \eqref{K0Kj transform} is different from (\ref{transformation A}), the transformation of the wave function should be also modified, namely
\be
\begin{cases}
K'_0=K_0 +\partial_0 g(x) \, ,\\
K'_j=K_j +4\partial_j g(x) \, ,\\
\Psi'=\Psi \exp\Bigl(\frac{im}{\hbar c} \int (\partial_0 g(x) dx^0 +4\partial_j g(x) dx^j)\Bigr) \, ,
\end{cases}
\label{combined transformation}
\ee
which leave the equation \eqref{gravitational wave equation} invariant. Indeed, according to \eqref{combined transformation}, we have
\bea
&&\mathscr{D}'_\mu \Psi'
=\partial_\mu \Psi' -\frac{im}{\hbar c} K'_\mu \Psi'
=(\mathscr{D}_\mu \Psi)
\exp\Bigl(\frac{im}{\hbar c} \int (\partial_0 g(x) dx^0 +4\partial_j g(x) dx^j)\Bigr) \, ,
\nonumber\\
&&(\mathscr{D}'_\mu)^2 \Psi'
=(\mathscr{D}^2_\mu \Psi)
\exp\Bigl(\frac{im}{\hbar c} \int (\partial_0 g(x) dx^0 +4\partial_j g(x) dx^j)\Bigr) \, .
\label{D square}
\eea
Finally, according to \eqref{D square}, 
\be
 -\frac{\hbar^2}{2m}\vec{\mathscr{D}}'^2 \Psi'
 - i\hbar c \mathscr{D}'_0 \Psi' =  \left( -\frac{\hbar^2}{2m}\vec{\mathscr{D}}^2 \Psi
 - i\hbar c \mathscr{D}_0 \Psi \right) e^{\frac{im}{\hbar c} \int (\partial_0 g(x) dx^0 +4\partial_j g(x) dx^j) }  \, .
\label{gravitational wave equation2}
\ee
Therefore, in general:
\be
   -\frac{\hbar^2}{2m}\vec{\mathscr{D}}^2 \Psi
 - i\hbar c \mathscr{D}_0 \Psi  = 0 
 \quad \Longrightarrow \quad 
 -\frac{\hbar^2}{2m}\vec{\mathscr{D}}'^2 \Psi'
 - i\hbar c \mathscr{D}'_0 \Psi' = 0 \, ,
\label{gravitational wave equation3}
\ee
which shows the invariance of  the Schr\"odinger equation \eqref{gravitational wave equation}.

\section{Gravitational Aharonov-Bohm effect\label{GAB effect}}

According to the solution \eqref{wave function}, the wave function acquires a gravitational phase factor of the form $e^{i\phi}$, where the phase $\phi$ is explicitly given by Eq.~\eqref{gravitational phase}. In close analogy with the electromagnetic case, this phase factor does not alter the physical content of the wave function when the particle is considered in isolation: the probability density and the expectation values of observables remain unchanged with respect to the free-particle case. In other words, the overall phase factor is unobservable in a single-path experiment. However, the situation changes fundamentally when interference between two wave functions is considered. In an interference experiment, the relative phase between the two components of the wave function becomes physically relevant, and the gravitational phase factor produces observable modifications of the interference pattern. This is precisely the mechanism underlying the Aharonov--Bohm effect, as originally pointed out in the electromagnetic context in Ref.~\cite{Aharonov:1959fk}.

To illustrate this point, let us consider a thought interference experiment of the type depicted in Fig.~\ref{cylindrical shell}. In close analogy with the original setup discussed in Ref.~\cite{Aharonov:1959fk}, the total wave function describing the superposition of the two alternative paths is given by
\begin{equation}
\Psi=\Psi_1 e^{i\phi_1} +\Psi_2 e^{i\phi_2} \, ,
\label{interference}
\end{equation}
where $\Psi_1$ and $\Psi_2$ denote the free wave functions associated with paths ``1'' and ``2'', respectively, while $\phi_1$ and $\phi_2$ are the corresponding gravitational phases acquired along each path. The physically observable quantity in such an interference experiment is the relative phase difference $\phi_2-\phi_1$, which encodes the influence of gravity on the interference pattern. If this phase difference is non-vanishing, the resulting interference fringes are shifted with respect to the free-particle case, thereby providing a direct signature of the gravitational interaction.

In this section, we analyze in detail the experiment illustrated in Fig.~\ref{cylindrical shell}. In particular, we will show that the interference pattern can be affected by a gravitomagnetic field confined in a region from which the particles are excluded, even though the particles propagate entirely in a region where the gravitomagnetic field vanishes. This behavior is completely analogous to the electromagnetic Aharonov--Bohm effect, where the magnetic field confined inside a solenoid influences the interference of electrons traveling in a field-free region, and it highlights the nonlocal character of the gravitational Aharonov--Bohm effect.

\subsection{The interference between two paths}

In Ref.~\cite{Chiao:2013ht}, an interference experiment involving two paths enclosing a ``solenoid'' that carries circulating mass currents was discussed within the framework of the Maxwell-like equations of gravity~\cite{Braginsky:1976rb}. The authors concluded that such mass currents give rise to a gravitational analogue of the vector Aharonov--Bohm effect. Inspired by their analysis, we investigate this effect here using a complementary approach based on the spacetime metric generated by the gravitational source. This method differs from the one employed in Ref.~\cite{Chiao:2013ht} and has the advantage of establishing a direct connection between the rotation of the shell and the resulting interference pattern. Our strategy is the following: we first determine the spacetime metric associated with the rotating source, then derive the corresponding gauge potential from the metric, and finally compute the gravitationally induced phase difference between the two interfering paths.

\begin{figure}[h]
\centering
\includegraphics[scale=0.2]{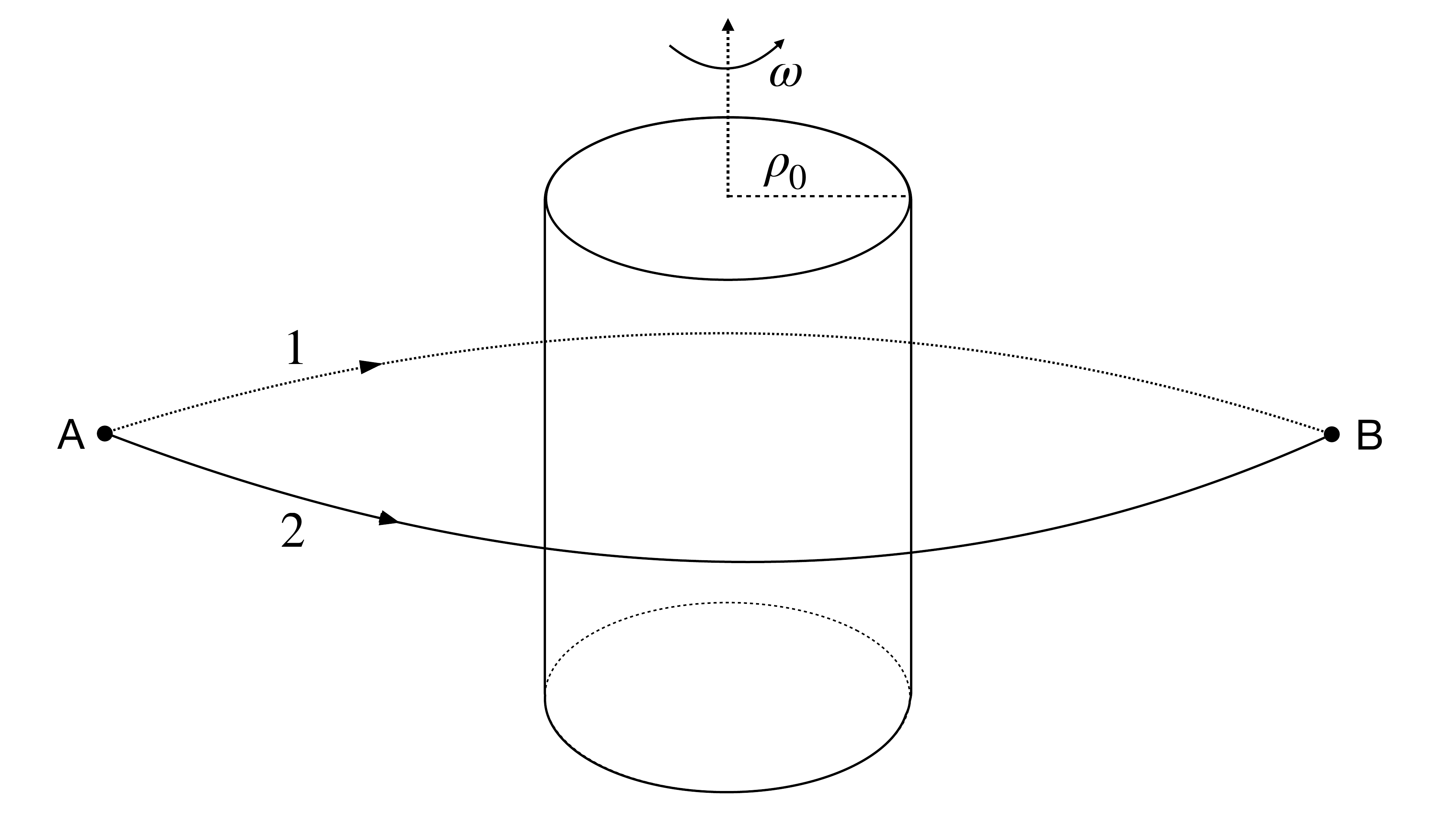}
\caption{The interference between two paths ``1'' and ``2''. The paths enclose an infinitely long and infinitely thin rotating shell which has a radius $\rho_0$ and an angular velocity $\omega$.}
\label{cylindrical shell}
\end{figure}

Specifically, we consider an interference experiment in which two paths enclose an infinitely long, infinitely thin rotating cylindrical shell of radius $\rho_0$, as illustrated in Fig.~\ref{cylindrical shell}. The cylindrical symmetry of the source allows for a tractable analytical treatment, while still capturing the essential physics of the gravitational vector Aharonov--Bohm effect.

The spacetime induced by this rotating shell, expressed in cylindrical coordinates $(ct,\rho,\varphi,z)$, is given by~\cite{Frolov1987}:
\begin{equation}
ds^2 =-\Bigl(1-\frac{a(\rho)}{2} \Bigr) (dx^0)^2
+\Bigl(1+\frac{a(\rho)}{2} \Bigr) (d\rho^2+\rho^2 d\varphi^2 +dz^2)
+2b(\rho) d\varphi dx^0 \, ,
\label{metric cylinder}
\end{equation}
where $x^0=ct$ and the functions $a(\rho)$ and $b(\rho)$ are defined as follows\footnote{The expressions \eqref{a rho} and \eqref{b rho} differ from equations (2.10) and (2.11) of Ref.~\cite{Frolov1987}. This discrepancy arises because the unit convention $G=1$ was adopted in Ref.~\cite{Frolov1987}, although this choice was not explicitly stated. Furthermore, we note that equations (2.1), (2.9), and (2.11) of Ref.~\cite{Frolov1987} should also be modified if $\omega$ is interpreted as the angular velocity. These modifications can be implemented by replacing $\omega$ with $\omega\rho_0$.}:
\begin{eqnarray}
a(\rho)&=&-\frac{8G\mu}{c^2} \Theta(\rho-\rho_0)\, {\rm ln}\frac{\rho}{\rho_0} \, ,
\label{a rho}
\\
b(\rho)&=&\frac{4G\mu}{c^3} \omega \rho_0^2 \Bigl[\frac{\rho^2}{\rho_0^2} \Theta(\rho_0-\rho) +\Theta(\rho-\rho_0) 
\Bigr] \, ,
\label{b rho}
\end{eqnarray}
where $\mu$ denotes the linear mass density of the cylindrical shell (for a finite shell of mass $M$ and length $L$, one has $\mu=M/L$), $\omega$ is the angular velocity of the shell, and $\Theta(T)$ is the unit step function defined by
\begin{equation}
\Theta(T)=
\begin{cases}
1, & \text{if $T\ge 0$} \, ,\\
0, &\text{if $T< 0$} \, .
\end{cases}
\end{equation}
From Eqs.~\eqref{a rho} and \eqref{b rho}, it is evident that the rotation of the shell is encoded entirely in the function $b(\rho)$, while the function $a(\rho)$ is independent of the angular velocity and describes the static gravitational field produced by the mass distribution of the shell.

It is important to remark that the metric \eqref{metric cylinder} is valid only in the weak-field region where the condition $|a(\rho)|\ll 1$ is satisfied~\cite{Frolov1987}. This requirement translates into the inequality
\begin{equation}
\Bigl|{\rm ln}\frac{\rho}{\rho_0} \Bigr|
\ll \frac{c^2}{8G\mu} \, , 
\qquad\text{for $\rho>\rho_0$} \, .
\label{a ll 1}
\end{equation}
Indeed, since $|a(\rho)|\rightarrow \infty$ as $\rho\rightarrow \infty$, the weak-field condition $|h_{\mu\nu}|\ll 1$ would be violated at sufficiently large distances, and the metric \eqref{metric cylinder} ceases to be reliable in that limit.

For the purpose of applying the metric \eqref{metric cylinder} to the interference experiment, it is convenient to rewrite it in Cartesian coordinates $(ct,x,y,z)$ by means of the standard coordinate transformation
\begin{equation}
g_{\mu\nu} =\frac{\partial x^{\mu'}}{\partial x^\mu} \frac{\partial x^{\nu'}}{\partial x^\nu} g_{\mu' \nu'} \, ,
\end{equation}
where the primed coordinates refer to the cylindrical system. A straightforward calculation yields
\begin{equation}
g_{\mu\nu}=
\begin{pmatrix}
\frac{a}{2}-1&-\frac{y}{\rho^2}b&\frac{x}{\rho^2}b&0\\
-\frac{y}{\rho^2}b&\frac{a}{2}+1&0&0\\
\frac{x}{\rho^2}b&0&\frac{a}{2}+1&0\\
0&0&0&\frac{a}{2}+1
\end{pmatrix},
\end{equation}
from which we immediately read off the perturbation $h_{\mu\nu}=g_{\mu\nu}-\eta_{\mu\nu}$:
\begin{equation}
h_{\mu\nu}=
\begin{pmatrix}
\frac{a}{2} &-\frac{y}{\rho^2}b&\frac{x}{\rho^2}b&0\\
-\frac{y}{\rho^2}b&\frac{a}{2} &0&0\\
\frac{x}{\rho^2}b&0&\frac{a}{2} &0\\
0&0&0&\frac{a}{2}
\end{pmatrix}.
\label{h mu nu}
\end{equation}

Before proceeding to the calculation of the gauge potential $A_\mu$, it is useful to compute the trace ${h^\alpha}_\alpha$. Using the metric signature $(-,+,+,+)$, we find
\begin{equation}
{h^\alpha}_\alpha =-h_{00} +\sum_j h_{jj}
=a \, .
\label{h}
\end{equation}
The potential $A_\mu$ is then obtained from Eq.~\eqref{Anu}, which gives
\begin{eqnarray}
A_0 &=&\frac{1}{2}c^2 \bar{h}_{00}
=\frac{1}{2}c^2\Bigl(h_{00} -\frac{1}{2} {h^\alpha}_\alpha \eta_{00}
\Bigr) \, ,
\nonumber\\
A_j &=& \frac{1}{2}c^2 \bar{h}_{j0}
=\frac{1}{2}c^2\Bigl(h_{j0} -\frac{1}{2} {h^\alpha}_\alpha \eta_{j0}
\Bigr)
=\frac{1}{2}c^2 h_{j0} \, .
\label{A bar h}
\end{eqnarray}
Substituting Eqs.~\eqref{h mu nu} and \eqref{h} into Eq.~\eqref{A bar h}, we obtain the explicit components of the gauge potential:
\begin{equation}
A_0 =\frac{a}{2} c^2  \, ,
\qquad
A_x= -\frac{y}{2\rho^2}b c^2 \, ,
\qquad
A_y= \frac{x}{2\rho^2}b c^2 \, ,
\qquad
A_z=0 \, .
\label{A components}
\end{equation}
Then, using the relation \eqref{relation KA} between $A_\mu$ and $K_\mu$, we arrive at
\begin{equation}
K_0 =\frac{a}{4}c^2 \, ,
\qquad
K_x= -\frac{y}{\rho^2}b c^2\, ,
\qquad
K_y= \frac{x}{\rho^2}b c^2\, ,
\qquad
K_z=0\, .
\label{K components}
\end{equation}

Substituting the components \eqref{K components} into the expression for the gravitational phase \eqref{gravitational phase}, the phase acquired by a particle propagating along a given path reads
\begin{equation}
\phi=\frac{m}{\hbar c}\int \Bigl[ \frac{a}{4}c^2 c dt +\frac{b}{\rho^2} c^2 (-ydx+xdy)\Bigr]
=\frac{mc}{\hbar}\Bigl( \frac{c}{4}\int_{t_0}^t a dt
+\int_{\varphi_0}^\varphi b d\varphi
\Bigr)\, .
\label{gravitational phase2}
\end{equation}
On the other hand, according to Eq.~\eqref{b rho}, the functions $a(\rho)$ and $b(\rho)$ take the explicit forms
\begin{equation}
a(\rho)=
\begin{cases}
0, &\text{for $\rho\le \rho_0$}\, , \\
-\frac{8G\mu}{c^2} {\rm ln}\frac{\rho}{\rho_0}, &\text{for $\rho>\rho_0$}\, ,
\end{cases}
\qquad
b(\rho)=
\begin{cases}
\frac{4G\mu}{c^3} \omega\rho^2, &\text{for $\rho\le \rho_0$}\, , \\
\frac{4G\mu}{c^3} \omega\rho_0^2, &\text{for $\rho>\rho_0$}\, .
\end{cases}
\label{ab in out}
\end{equation}
Therefore, for a path lying entirely outside the cylindrical shell, the gravitational phase \eqref{gravitational phase2} simplifies to
\begin{equation}
\phi
=\frac{m}{\hbar}\Bigl[ -2G\mu \int_{t_0}^t {\rm ln}\frac{\rho(t)}{\rho_0} dt
+\frac{4G\mu}{c^2} \omega\rho_0^2(\varphi-\varphi_0)
\Bigr] \, .
\label{gravitational phase out}
\end{equation}

Let us now consider the interference between two paths that both lie outside the cylindrical shell and together enclose it. The gravitational phase difference between the two paths is given by
\begin{equation}
\Delta\phi
=\phi_2-\phi_1
=-\frac{2Gm\mu}{\hbar}\Bigl[ \Bigl(\int_{t_0}^t {\rm ln}\frac{\rho(t)}{\rho_0} dt\Bigr)_2-\Bigl(\int_{t_0}^t {\rm ln}\frac{\rho(t)}{\rho_0} dt\Bigr)_1
\Bigr]
+\frac{8\pi G m \mu}{\hbar c^2} \omega\rho_0^2 \, ,
\label{phase difference out}
\end{equation}
where the subscripts ``1'' and ``2'' refer to the first and second paths, respectively, and Eq.~\eqref{gravitational phase out} has been used. The second term on the right-hand side of Eq.~\eqref{phase difference out} is independent of the specific geometry of the paths and depends only on the angular velocity of the shell and on its radius. This term represents the gravitational analogue of the Aharonov--Bohm phase shift, arising from the rotation of the source even though the particles never enter the region where the gravitomagnetic field is non-vanishing.

\subsection{Discussion\label{discussion}}

The expression for the phase difference in Eq.~\eqref{phase difference out} consists of two distinct contributions. The first term, involving two path integrals, depends explicitly on the geometry of the trajectories and is therefore path-dependent. In contrast, the last term is independent of the specific shape of the paths and depends only on global properties of the configuration, namely the angular velocity and the radius of the rotating shell. For this reason, the last term can be regarded as a topological invariant, and it clearly exhibits the non-local character of the gravitational Aharonov--Bohm effect in the interference experiment illustrated in Fig.~\ref{cylindrical shell}.

The last term in Eq.~\eqref{phase difference out} also demonstrates that the interference pattern is affected by the rotation of the shell. This effect has no classical counterpart, since in the region outside the shell the geodesic equation
\begin{equation}
\frac{d^2 x^\mu}{ds^2} +\Gamma^\mu_{\alpha\beta}\frac{dx^\alpha}{ds}\frac{dx^\beta}{ds} =0
\end{equation}
is completely independent of the rotation of the shell. Indeed, according to Eq.~\eqref{ab in out}, in the exterior region the function $b(\rho)$ is constant with respect to the coordinates, and therefore it gives no contribution to the connection coefficients $\Gamma^\mu_{\alpha\beta}$. Consequently, the angular velocity $\omega$ does not appear in the geodesic equation, and no classical ``force'' due to the rotation acts on a particle propagating outside the shell. Nevertheless, as shown by the last term in Eq.~\eqref{phase difference out}, the rotation of the shell does influence the quantum mechanical phase and hence the interference pattern. This is therefore a purely quantum mechanical effect, closely analogous to the electromagnetic Aharonov--Bohm effect. It is worth mentioning that a related effect, albeit in a different physical setting, was identified in Ref.~\cite{Frolov1987}, where the authors studied the energy levels of a massive scalar particle confined in a cylindrical cavity enclosing the rotating shell. Using the linear approximation for the gravitational field, they showed that the rotation of the shell contributes to the energy spectrum of the particle, even though the rotation does not affect the Riemann curvature tensor in the exterior region.

A practical way to detect this rotation-induced term consists in reversing the direction of rotation of the shell, i.e., replacing $\omega$ with $-\omega$, and measuring the resulting shift of the interference fringes. As argued above, the rotation does not affect the classical motion of the particles outside the shell, so it is reasonable to assume that the paths in Eq.~\eqref{phase difference out} remain unchanged when the rotation is reversed. Under this assumption, the first term in Eq.~\eqref{phase difference out} is also unchanged, while the last term simply changes sign. The fringe shift is therefore given by
\begin{equation}
N=\frac{|\Delta\phi' -\Delta\phi |}{2\pi}
=\frac{8 G m \mu}{\hbar c^2} \omega\rho_0^2 \, .
\end{equation}

Finally, we provide a physical interpretation of the effect manifested in Eq.~\eqref{phase difference out} in close analogy with the electromagnetic Aharonov--Bohm effect. To this end, we first determine the gravitoelectromagnetic fields. Substituting Eq.~\eqref{A components} into Eq.~\eqref{G0 nu rho}, we obtain
\begin{eqnarray}
F_{\nu\rho} 
&=& c^2 \begin{pmatrix}
0& -\frac{1}{2}\partial_x a &  -\frac{1}{2}\partial_y a &0\\
 \frac{1}{2}\partial_x a & 0 & \frac{1}{2}\bigl[\partial_x \bigl(\frac{xb}{\rho^2} \bigr)+ \partial_y \bigl(\frac{yb}{\rho^2} \bigr)\bigr] &0\\
\frac{1}{2}\partial_y a & -\frac{1}{2}\bigl[\partial_x \bigl(\frac{xb}{\rho^2} \bigr)+ \partial_y \bigl(\frac{yb}{\rho^2} \bigr)\bigr] &0& 0 \\
0&0&0&0
\end{pmatrix},
\label{G0 cartesian}
\\
E_x&=&\frac{c^2}{2}\partial_x a \, ,
\qquad
E_y=\frac{c^2}{2}\partial_y a \, ,
\qquad
E_z=0 \, ,
\label{E expressions}
\\
B_x&=&0 \, ,
\qquad
B_y=0 \, ,
\qquad
B_z=\frac{c^2}{2}\Bigl[\partial_x \Bigl(\frac{xb}{\rho^2}\Bigr)+\partial_y \Bigl(\frac{yb}{\rho^2}\Bigr)\Bigr] \, .
\label{B expressions}
\end{eqnarray}
Using the explicit expressions for $a(\rho)$ and $b(\rho)$ given in Eqs.~\eqref{a rho} and \eqref{b rho}, the components $E_x$, $E_y$, and $B_z$ can be written as
\begin{eqnarray}
E_x&=&-\frac{4G\mu x}{\rho}\Bigl[
\delta(\rho-\rho_0) {\rm ln}\frac{\rho}{\rho_0}
+\frac{1}{\rho}\Theta(\rho-\rho_0)
\Bigr]
=\begin{cases}
0 \, , &\text{for $\rho< \rho_0$} \, , \\
-\frac{4G\mu x}{\rho^2} \, , &\text{for $\rho\ge \rho_0$} \, , 
\end{cases}
\label{Ex}
\\
E_y&=&-\frac{4G\mu y}{\rho}\Bigl[
\delta(\rho-\rho_0) {\rm ln}\frac{\rho}{\rho_0}
+\frac{1}{\rho}\Theta(\rho-\rho_0)
\Bigr]
=\begin{cases}
0 \, , &\text{for $\rho< \rho_0$} \, , \\
-\frac{4G\mu y}{\rho^2} \, , &\text{for $\rho\ge \rho_0$} \, , 
\end{cases}
\label{Ey}\\
B_z&=&\frac{2G\mu\omega\rho_0^2}{\rho c}\Bigl[\frac{2\rho}{\rho_0^2}\Theta(\rho_0-\rho) +\Bigl(1-\frac{\rho^2}{\rho_0^2}\Bigr)\delta(\rho-\rho_0)
\Bigr]
=
\begin{cases}
\frac{4G\mu\omega}{c} \, , &\text{for $\rho\le \rho_0$} \, , \\
0 \, , &\text{for $\rho> \rho_0$} \, .
\end{cases}
\label{Bz}
\end{eqnarray}

We now summarize the main features of the gravitoelectromagnetic fields in Eqs.~\eqref{E expressions} and \eqref{B expressions} (see also Eqs.~\eqref{Ex}, \eqref{Ey}, and \eqref{Bz}) as follows: (I) The rotation of the shell manifests itself exclusively in the gravitomagnetic field, while the gravitoelectric field is entirely determined by the static mass distribution and is independent of $\omega$. This is a direct consequence of the definitions of the functions $a(\rho)$ and $b(\rho)$ in Eqs.~\eqref{a rho} and \eqref{b rho}. (II) The gravitomagnetic field vanishes identically in the exterior region, and inside the shell it has only the $z$-component, given by $B_z=4G\mu\omega/c$. The characteristic (II) is completely analogous to the magnetic field configuration of an infinitely long solenoid carrying a steady current~\cite{Griffiths}.

We now demonstrate that the gravitomagnetic field confined inside the shell is precisely responsible for the last term in the phase difference \eqref{phase difference out}. In close analogy with the electromagnetic case discussed in Ref.~\cite{Aharonov:1959fk}, the phase difference \eqref{phase difference out} can be expressed as the line integral of the gauge field along a closed spacetime path,
\begin{equation}
\Delta\phi=\frac{m}{\hbar c}\oint_C K_\mu dx^\mu .
\label{gravitational phase close}
\end{equation}
According to Eq.~\eqref{K components}, the rotation of the shell appears only in the spatial components of $K_\mu$. Therefore, the contribution to the phase difference \eqref{gravitational phase close} arising from the rotation is given by
\begin{eqnarray}
\Delta\phi_{\rm r}
&=&\frac{m}{\hbar c}\oint_L \vec{K} \cdot d\vec{x}
\nonumber\\
&=&\frac{m}{\hbar c}\iint_S \nabla\times\vec{K}\cdot d\vec{S}
\nonumber\\
&=&\frac{2m}{\hbar c}\iint_S \nabla\times\vec{A}\cdot d\vec{S}
\nonumber\\
&=&\frac{2m}{\hbar c}\iint_S \vec{B}\cdot d\vec{S}
\nonumber\\
&=&\frac{2m}{\hbar c}\Phi \, ,
\label{phase difference space}
\end{eqnarray}
where $\Phi$ is the ``gravitomagnetic flux'' enclosed by the path, in complete analogy with the magnetic flux of electromagnetism. In deriving Eq.~\eqref{phase difference space}, we have used the relation \eqref{relation KA} and the last equation in \eqref{G0 nu rho}. Equation \eqref{phase difference space} shows explicitly that $\Delta\phi_{\rm r}$ is determined by the gravitomagnetic field inside the shell. Let us quickly check whether \eqref{phase difference space} correctly produces the last term of \eqref{phase difference out}. Since the gravitomagnetic field inside the shell has only the $z$-component given by Eq.~\eqref{Bz}, we have
\begin{equation}
\frac{2m}{\hbar c}\Phi
=\frac{2m}{\hbar c} \times \frac{4G\mu\omega}{c} \times \pi \rho_0^2
=\frac{8\pi Gm\mu}{\hbar c^2}\omega \rho_0^2 \, ,
\end{equation}
which coincides exactly with the last term in Eq.~\eqref{phase difference out}.

The above analysis of the interference experiment in Fig.~\ref{cylindrical shell} shows that the non-vanishing gravitomagnetic field inside the rotating shell can affect the interference of two paths that enclose the shell while remaining entirely in the exterior region, where the gravitomagnetic field is zero. This phenomenon is closely analogous to the electromagnetic Aharonov--Bohm effect for a solenoid discussed in Ref.~\cite{Aharonov:1959fk}, and it can therefore be interpreted as a gravitational Aharonov--Bohm effect.

\section{Conclusion}

In this paper we have studied a gravitational Aharonov--Bohm effect in the spacetime of an infinitely long, infinitely thin rotating cylindrical shell. The gravitational field was assumed to be sufficiently weak that the formalism of gravitoelectromagnetism (GEM) could be applied. Within this framework, we defined a gauge potential and derived a Schr\"odinger equation describing the dynamics of a non-relativistic quantum particle in the presence of the weak gravitational field. The solution of the Schr\"odinger equation exhibits a gravitational phase factor that is directly related to the gauge potential. Based on this phase factor, we then analyzed the interference between two paths that enclose the rotating shell while remaining entirely in the exterior region.

Using the spacetime metric generated by the shell, we computed the gravitational phase difference between these two paths. A key result of our analysis is that the rotation of the shell contributes to the phase difference, even though this rotation has no influence on the classical motion of particles propagating outside the shell. In addition, we calculated the gravitoelectromagnetic fields induced by the shell. Our results show that the rotation contributes exclusively to the gravitomagnetic field, while the gravitoelectric field is determined solely by the static mass distribution and remains independent of the angular velocity. Outside the shell, the gravitomagnetic field vanishes identically, whereas inside the shell it has only the $z$-component. Consequently, in complete analogy with the magnetic field of a solenoid carrying a steady current, the phase difference is governed only by the ``gravitomagnetic flux,'' which is determined by the gravitomagnetic field confined inside the shell. The analysis of this gedanken experiment demonstrates that a gravitomagnetic field localized in one region can affect the quantum interference of particles propagating in a different region where the gravitomagnetic field is zero. This phenomenon is intrinsically quantum mechanical and can be regarded as a gravitational analogue of the Aharonov--Bohm effect.

As a thought experiment, let us imagine such a rotating shell placed somewhere in the universe. The spacetime metric \eqref{metric cylinder} induced by the shell reveals an interesting aspect concerning the detection of its rotation. The contribution of the rotation appears only in the components $g_{0i}$, which are independent of the coordinates in the exterior region. Therefore, as discussed in Sec.~\ref{discussion}, the rotation of the shell does not affect the geodesics of particles outside it. This implies that the rotation cannot be detected through the scattering of test particles. It also cannot be detected by means of the gravitational redshift, since the latter depends only on the $g_{00}$ component of the metric~\cite{Landau:1975pou}. Nevertheless, as we have shown in this paper, the rotation does affect the interference of two coherent beams that enclose the shell. This situation is somewhat reminiscent of dark matter, which can be detected gravitationally but not optically, although the analogy is not exact.

Finally, we outline some potential extensions of the present work. Since the gravitational source considered in the interference experiment of Fig.~\ref{cylindrical shell} is an idealized rotating cylindrical shell, it would be worthwhile to investigate whether a gravitational Aharonov--Bohm effect also arises in the spacetime of more realistic, non-idealized objects. Whether the effect persists in such cases is a question left for future study. Furthermore, in this paper the gravitational field was assumed to be weak, and a natural extension would be to consider the strong-field regime. In that case, however, the GEM formalism may no longer be applicable, and new theoretical tools would likely be required to analyze the gravitational Aharonov--Bohm effect. Finally, an interesting direction for future work is the study of interference between paths enclosing a rotating cosmic string. In Ref.~\cite{Jensen:1993cx}, quantum systems in the exterior region of a rotating cosmic gauge string were investigated, and it was found that the angular momentum of a particle is shifted by a constant related to the angular momentum density of the string, even though the Riemann curvature vanishes in that region. This scenario bears a close resemblance to the gravitational Aharonov--Bohm effect discussed here, where the rotation of the shell gives a non-vanishing contribution to the gravitational phase difference.

Last comment concerns about quantum gravity \cite{Bambi:2023jiz, Modesto:2011kw}. Indeed, It is often stated that there is currently no direct experimental evidence for quantum gravity. However, as we have emphasized, neutron interference experiments in the presence of a gravitational potential already provided clear evidence for quantum effects in a gravitational context \cite{Colella:1975dq}. In those experiments, the dynamics of neutrons was governed by a Schr\"odinger equation in which the gravitational potential appeared as an operator acting on the wave function. This meant that the $h_{00}$ component of the metric perturbation was already treated quantum mechanically, even though the spatial components $h_{ij}$ (with $i,j=1,2,3$), which describe gravitational waves, were not quantized in that setting. The resulting theoretical predictions were in excellent agreement with experimental observations, and the situation was perfectly analogous to the treatment of the electromagnetic potential in quantum electrodynamics, where the scalar potential is quantized while the radiation field may be treated classically in certain approximations.

In the present paper, we went a step further by considering the quantization of the $h_{0i}$ components of the metric perturbation. Within the framework of gravitoelectromagnetism, these components are directly related to the gravitomagnetic vector potential, in close analogy with the vector potential of electromagnetism. We investigated whether the quantization of $h_{0i}$ could lead to observable consequences and proposed a possible experimental setting in which such effects might, in principle, be tested. In this way, we aimed to extend the well-established quantum gravitational phenomenology associated with the scalar part of the metric to the vector part, thereby providing new insights into the quantum aspects of gravity.





\end{document}